\documentclass[sigconf]{acmart}

\AtBeginDocument{%
  \providecommand\BibTeX{{%
    \normalfont B\kern-0.5em{\scshape i\kern-0.25em b}\kern-0.8em\TeX}}}

\begin{document}

\title{Validation of a simulation model for FaaS performance benchmarking using predictive validation}

\author{David Ferreira Quaresma}
\authornotemark[1]
\affiliation{%
  \institution{Federal University of Campina Grande}
  \city{Campina Grande}
  \state{Paraíba}
  \country{Brazil}
}
\email{david.quaresma@ccc.ufcg.edu.br}

\author{Thiago Emmanuel Pereira}
\affiliation{%
  \institution{Federal University of Campina Grande}
  \city{Campina Grande}
  \state{Paraíba}
  \country{Brazil}}
\email{temmanuel@computacao.ufcg.edu.b}

\author{Daniel Fireman}
\affiliation{%
  \institution{Federal Institute of Alagoas}
  \city{Arapiraca}
  \state{Alagoas}
  \country{Brazil}}
\email{daniel.fireman@ifal.edu.br}

\renewcommand{\shortauthors}{Quaresma, Pereira and Fireman, et al.}

\begin{abstract}
In the paper Controlling Garbage Collection and Request Admission to Improve Performance of FaaS Applications\cite{quaresma-sbacpad}, we verified and evaluated the impact of memory management mechanics of programming languages in the context of Functions as a Service (FaaS) via simulation experiments. The results of this study pointed to an impact of up to $11.68\%$ on the response time of requests when a garbage collector procedure was executed during the execution of a CPU-bound function. As future work, we listed a few threats to the validity of the results attained, and among them, we cited the validation of the simulation model used. The validation of the model is important because it validates the results generated in the simulation experiments, which ensures realistic results. In this work, we proposed and executed a validation to the simulation model used in the previous work. To do so, we run measurement experiments in a public FaaS platform and simulation experiments of the same scenarios using the same simulator of the previous paper. Then, we validate the simulator by comparing the results obtained in both experiments to ensure that the simulation result and the measurement one are equivalent.
\end{abstract}

\maketitle

\section{Introduction}
\label{sec:introduction}
Function as a Service (FaaS) is a recent offer of computing as a service and a model of Serverless computing~\cite{A-Review-of-Serverless-Frameworks}. It is an event-driven computing model that is based on the execution of ephemeral stateless functions in containers~\cite{redhat:what-is-faas}. A function, in this model, is triggered by events predefined, and all the burden of resource management is outsourced to the platform. An event, for instance, might be a function invocation request from a user client or another function instance. In this way, FaaS promises to reduce the time to deliver software, as well as providing flexible scalability (on function level) and a more precise billing system based on the time in which each function has been processing events~\cite{Kumar2019,marcus:sbrc2019:serverless}.

Although popular and promising~\cite{infoq-serverless-popularity}, FaaS is still a new Cloud Computing offer. There are studies categorizing it~\cite{SPEC-RG:reference-for-faas}, evaluating its performance \cite{Evaluation-of-Production-Serverless-Computing-Environments, An-Evaluation-of-Open-Source-Serverless-Computing-Frameworks-Support-at-the-Edge}, and proposing new approaches to deal with already known problems~\cite{feitosa-middleware}. Still, there are open problems in this field and among them, we emphasize the performance degradation of FaaS platforms (and applications) due to the runtime background activity~\cite{quaresma-sbacpad}.

Using simulation,  we estimated the consequences of using Garbage Collector (GC) for FaaS scenarios~\cite{quaresma-sbacpad}. We observed that when a GC routine was triggered during function execution, the response time of this function call was increased by up to 11.68\%~\cite{quaresma-sbacpad}. We also proposed an approach to mitigate this impact by controlling garbage collection and request admission using the Garbage Collector Control Interceptor (GCI), a technique used to avoid the impact of GCs~\cite{FiremanSBRC2017, fireman-cloudcom}. The results attained of using the GCI in the FaaS context were the reduction of the microservice response time of up to 10.86\% and saving of resources costs of 7\%~\cite{quaresma-sbacpad}.

The previous study mentioned is still a work in progress. There are some threats to the validity of results that we made sure to point out, such as the lack of variety of used functions or a more representative set of FaaS functions, the usage of a synthetic workload upon a discrete probability distribution (Poisson), and the absence of validation for the simulator used. Although the simulator was properly tested and verified, no further validation was done. The validation process is important because it gives us credibility about its results. Thus, the main objective of this work is to validate the simulator used in the previous work and evaluate if the model and production environment are similar. To do so, we executed experiments using the simulation model and the scenarios of the previous study and compared them under statistical analysis.

The remainder of this paper is organized as follows. In Section~\ref{sec:background}, we introduce the basics of our previous work~\cite{quaresma-sbacpad}, such as the results and conclusions attained, and then we talk about main concepts of validation, such as validation terminology and validation approach to be used in this work. Then, in Section~\ref{sec:methodology}, we present our approach to validation, describing the simulator model and its components, the validation process used, and the measurements experiments. in Section~\ref{sec:results}, we present and discuss the results attained. Finally, in Section~\ref{sec:conclusion}, we present our final remarks and possible future work.
\section{Background}
\label{sec:background}
In this section, we describe the previous study we are going to continue as well as the basic concept behind this work, which is the validation process. In Section~\ref{subsec:previous-work} we introduce our previous study~\cite{quaresma-sbacpad} discussing what it is, what results were accomplished, and briefly how it was attained. Following, in Section~\ref{subsec:simulation-model-validation} we present the basics of the validation process we are going to use in this work, such as definition and general usage.

\subsection{Previous Work}
\label{subsec:previous-work}
As mentioned in the Introduction, in the previous work~\cite{quaresma-sbacpad}, we evaluated the drawback of letting FaaS functions be vulnerable to Garbage Collector (GC) activity. The paper “Controlling Garbage Collection and Request Admission to Improve Performance of FaaS Applications”~\cite{quaresma-sbacpad} not only showed that a GC might increase the response time by up to $11.68\%$, but also provided a solution to mitigate this impact that reduced the same metric of up to $10.86\%$ and reduced the resource costs by $7\%$. As said, although the results attained are very promising, some threats to its validity require further work. 

The previous work~\cite{quaresma-sbacpad} is a study that evaluated the relation of performance (in terms of service time) degradation of FaaS functions and the memory management procedures of the programming language in which the function was coded with. The main two scenarios studied were, initially, a function execution that was impacted by garbage collections and another function execution free of any GC intervention. In both scenarios, the function used was the image resizer, a popular FaaS function in the literature~\cite{oakes2018sock, akkus2018sand}.

Once the GC impact was properly evaluated with a workload much simpler, simulation experiments were performed to evaluate it under a more complex workload configuration. The complex workload configuration used was a series of function invocations in which the time between each invocation was defined following a probability distribution - the Poisson distribution to be precise. The FaaS platform behavior and function environment considered to the experiments were based on the AWS Lambda. The same process was executed to compare the feasibility of the solution proposed.

The results attained in this study, and its conclusions, were obtained using simulation experiments. The simulation model used is described further in Section~\ref{subsec:simulation-model}. For the scope of the previous work, the precision of the simulated distribution shape of the response time was essential since the strategy behind the solution suggested is highly impacted by the shape of the distribution.

\subsection{Model Validation}
\label{subsec:simulation-model-validation}
A key concept to understand this work is validation. While verification often stands for ensuring that a conceptual model and its implementation are correct~\cite{Validation-And-Verification-Of-Simulation-Models, about-validation-and-simulation-concepts}, model validation stands for a satisfactory accuracy of the computerized model with the intended application of the model within its domain of applicability~\cite{Terminology-for-model-credibility, about-validation-and-simulation-concepts}. In other words, validation is a process where the used model is tested to ensure that it does represent what it is proposed to represent, considering the scenarios that are important to it. It is an important process because it validates the model tested and also any information acquired from using this model in some simulation experiment.

There are a variety of methods used to validate simulation models and assert its validity, ranging from approaches such as comparisons between models to the usage of data generated by an actual system~\cite{mitre-validation}. In this study, to validate the simulator, we use a Predictive Validation technique. The Predictive Validation approach consists of using a simulation model to predict, or forecast, the behavior of the target system, and then compare the results attained to determine if they are the similar~\cite{validation-of-simulation-models}. We decided to use the Predictive Validation because it was the technique that best feat in out conditions, but there were two other interesting approaches as well, such as the Face Validity and the Comparison to Other Models.


\section{Methodology}
\label{sec:methodology}
The main goal of this work is to validate whether the simulator used in the previous work~\cite{quaresma-sbacpad} is able to produce realistic results from various scenarios. Thus, in this work, we perform a validation of the simulator mentioned using the predictive validation approach, as highlighted in Section~\ref{subsec:simulation-model-validation}. We evaluate if the results attained with the simulator are realistic by comparing them to measurement experiment results, which are, by default, realistic since they are collected from a real function in production. 

To avoid misleading understandings, we emphasize that since the input of the simulator is generated by measurement experiments, there are two sorts of measurement experiments. One of them (described in Section~\ref{subsec:simulation-inputs}) was used as simulation input, while the other one (described in Section~\ref{subsec:measurements-for-validation}) was used for comparison purposes. Thus, we will call the measurement experiments used for the simulation input just as input experiments.

The remainder of this section is described as follows: In Section~\ref{subsec:simulation-model}, we discuss the simulation model used in the previous~\cite{quaresma-sbacpad} and current work and briefly describe each main component of the Simulator. Then, in Section~\ref{subsec:validation-process}, we discuss the validation process, metrics, and the scope of this validation. Finally, in Section~\ref{subsec:measurement-experiments} we describe the measurement experiments performed to provide input to the simulations as well as realistic results for comparison and validation purposes.

\subsection{Simulation Model}
\label{subsec:simulation-model}
Even though lots of infrastructures of public FaaS platforms have many aspects in common, such as a simple deployment, a dynamic resource provisioning, and a pay-per-use billing, they are not equals. In fact, they vary a lot in many other aspects. For instance, GCP Cloud Functions~\cite{google-cloud-functions-page} and AWS Lambda~\cite{aws-function-page} differ in terms of available events that triggers the execution of functions, resource specs, and supported execution environment. Not only the platforms differ from each other, but applications (the functions) might as well, and their workload too (which veries from applications to applications). 

So, during the modeling of the simulator used, some challenges showed up. First, since FaaS is fairly new, how to choose a proper function, platform, and workload to representatively evaluate it? and, for instance, even taking a public FaaS provider as a reference, how to properly simulate its environment if its internal settings are typically not public? To address these issues, in~\cite{quaresma-sbacpad}, we decided to use AWS Lambda platform as a reference and executed a couple of functions on it to collect as much internal information as possible. Regarding a representative function, we solely choose to use as reference a FaaS function popular in literature and run it using the settings we were able to retrieve. Finally, regarding to the workload, to make things simple, we choose a synthetic workload using the Poisson distribution to generate the inter-arrival of each function invocation during simulations. Thus, the model to validate is purposed to simulate the AWS Lambda environment, by a chosen function, and a chosen workload. It is a simplified version of the platform having only the main abstractions of it.

The decision of use the AWS Lambda was made based on the fact that AWS Lambda platorm is a very popular public offering in this field~\cite{awsinsider-was-lambda-popularity,Top-4-Serverless}. Among its characteristics, we highly considered and applied in this simulation model the serial execution of requests by function replicas~\cite{aws:lambda-concurrency-level}, the termination of function replicas after an idle period of time~\cite{A-Cloud-Guru-about-idleness}, and the reuse of replicas most used recently. The internal settings collected were used to understand the platform and also to setup the measurement experiment (read Section~\ref{subsec:measurement-experiments} for further details about the measurement experiments) since in our study we used the same internal settings of AWS Lambda to run experiments and measure the response time to be used as blueprint by the simulations. Lastly, the chosen function, as introduced in~\ref{sec:background}, was a image resizer function popular in FaaS literature~\cite{oakes2018sock, akkus2018sand}.

\begin{figure}[ht]
  \centering
  \includegraphics[width=1\linewidth]{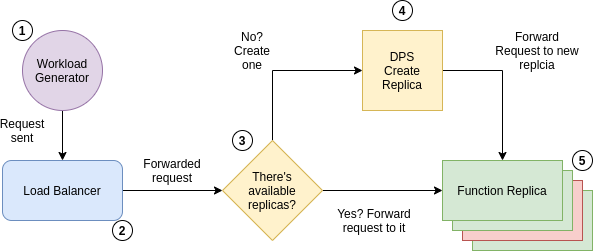}  
\caption{Request's flow through the simulated model components. The Workload Generator (WG) periodically sends requests to the Load Balancer (LB), which redirects it to the Dynamic Resources Provisioning System (DRPS), the one that make sure to redirect the request to a available function, even if it have to create another function replica.}
\label{fig:model}
\end{figure}

This simulation model is able to simulate a FaaS platform environment by only receiving a specific input - a list of response time of each function invocation measured from a FaaS function deployed. This list must be collected through measurement experiments using a sequential workload (as presented later in~\ref{subsec:simulation-inputs}) to avoid functions invocations of running concurrently, and is required one to each function replica to be simulated. Thus, the simulator reproduce, to each function replica, the respective input file and simulate how the provider scale up and/or scale down based on the demand. Whenever a new invocation arrives and there is no available replica, in this simulation model a new function replica is created and the response time is impacted by a cold start.

The simulator model~\cite{github:gci-simulator} is composed of four main components~\cite{quaresma-sbacpad}: 1) the Workload Generator (WG) - responsible to generate events processed by the function replicas according to a given statistical distribution; 2) the Load Balancer (LB) - responsible to balance load among replicas in a given algorithm fashion; 3) a Dynamic Resources Provisioning System (DRPS) - responsible for scaling the pool of replicas to match the demand; and 4) the Function Replicas - the one responsible to handle requests. Figure \ref{fig:model} summarizes the request processing flow and the relationship between the components of the model. We detail these components following below.

\subsubsection{Workload Generator}
\label{subsec:workload}

The first entity in this model is the Workload Generator (WG). The primary goal of WG is to send requests to the load balancer according to a statistical distribution. More precisely, the WG defines the time between the sending of each request based on a discrete probability distribution. So, after a given number of simulated time, the WG sends to the load balancer a function invocation. In this work, we modeled the inter-arrival time according to a Poisson distribution. In Figure \ref{fig:model}, it is the first step of the requests flow in the simulation environment. 

\subsubsection{Load Balancer}
\label{subsec:balancing}

The second entity of the simulated model is the Load Balancer (LB). We chose this model based on observations of how AWS Lambda works. Since we know AWS Lambda scale down its replicas based on inactivity time, we judge that its LB cannot distribute the load using popular policies such as Round-Robin, once it would uniformly restart the idle time counter. So, our LB receives all generated requests from the WG and schedules them to an available replica. Regarding the scheduling, the LB chooses the replica which has most recently become available. Lastly, the requests are processed according to the first-in-first-out (FIFO) policy. In Figure \ref{fig:model}, it is the second component an invocation hits after being generated by the WG. 

\subsubsection{Dynamic Resources Provisioning System}
\label{subsec:dynamic_prov}

The third component of this simulator model is the Dynamic Resources Provisioning System (DRPS). It is responsible to start new function replicas or terminating the idle ones whenever is needed. It starts new replicas when a new request arrives, and there is no available replica to process it. This is the case when all replicas are busy processing other requests or when there is no replica available. The DRPS automatically terminates replicas when they remain idle for some time. This idleness period until replica removal is configurable, and the default value is $5$ minutes. In Figure \ref{fig:model}, it is composed by the steps 3 and 4, and is the last component an invocation hits before being received by a function replica.

\subsubsection{Function Replica}
\label{sec:instance}

The last component of this model is the Function Replica. Our model of the Function Replica dictates its performance, as response times, when processing requests. In addition to the response time, the modeled response time also accounts for runtime environment interference, for instance, the impact of runtime startup, automatic memory management, and cold starts. This last one, cold starts, is a problem of virtualization technique that FaaS has inherited due to the fact that functions are executed in containers~\cite{coldstartref}. The cold start a is delay in the function execution due to the fact that there is no ready resource, or container, to host the execution~\cite{feitosa-middleware}.

The model of each function replica is represented as a sequence of \textit{(duration, status code)} tuples. Each tuple of this sequence represents the processing of a request. Thus, the model simulates the replica behavior by reproducing the duration and the status code of each tuple in that sequence. For instance, upon receiving the first request, the simulated replica returns the first tuple of the set associated with it. The procedure repeats to all the following replicas and requests.

In Figure \ref{fig:model}, when the generated function invocation hits the step 5, the Function Replica assign to this invocation the duration and status code of the current tuple, simulating the behavior of this invocation to be the same of one of the invocations measured before. In the Figure, function might green or red to illustrates that it might be available or busy. To better represent the state of practice, in our simulations, the model of the function replicas was instantiated based on experimental measurements (explained in Section~\ref{subsec:measurement-experiments}), and each simulated replica is associated with a result of a different measurement experiment.

\subsection{Validation Process}
\label{subsec:validation-process}

\begin{figure}[ht]
  \centering
  \includegraphics[width=1\linewidth]{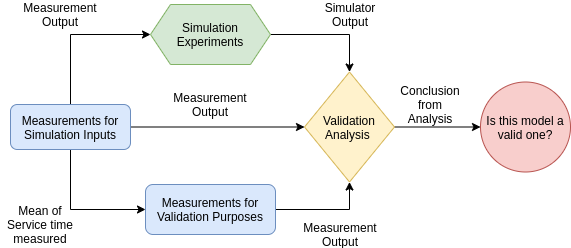}  
\caption{Flow diagram of the validation process. The first step was the measurement experiment to collect the simulation input and statistics from the function deployed. Then a simulation and measurement experiment were executed based on this data. Finally, the whole data was analyzed in order to validate the simulation model}
\label{fig:validation-process}
\end{figure}

The process of validation used in this work was predictive validation. It consists of executing measurement and simulation experiments, both for the same scenarios, and comparing the results obtained under statistical analysis to assert if the simulation model was able to forecast the system behavior. This section describes in detail the validation process designed and each step of its execution as shown in Figure~\ref{fig:validation-process}. Later, we discuss the simulation experiments used for this process.  

To perform the simulation experiment we need to generate the proper input to the simulator, which we get from the input experiments pointed in Section~\ref{sec:instance} and described further in Section~\ref{subsec:simulation-inputs}. Briefly, the input experiments is a simple measurement experiments, an experiment using a sequential workload in order to avoid concurrent execution of functions. After collecting the input required, we execute simulation and measurement experiments using a more complex workload for the same scenarios. Once the measurement and simulation experiment are done, we compare, as said, its results under a confidence interval to ensure that the simulation results can be representative and relevant. We want to validate the results of the simulator and check whether its results are realistic for the scenarios studied previously~\cite{quaresma-sbacpad}.

Figure~\ref{fig:validation-process} illustrates the whole process of validation in a flow diagram. The first step in the validation process is the input experiments aimed to collect the simulation input and statistical measures from the function behavior in the production environment. More details about this experiments in~\ref{subsec:simulation-inputs}. Next, we execute two sorts of experiments: a simulation experiment (described in Section~\ref{subsec:simulation-experiments}) and a measurement experiment for validation purposes (described in Section~\ref{subsec:measurements-for-validation}). Both these measurement and simulation experiments focus on collect mainly the service time, i.e. the time took to execute function code.

After that, as in the Figure~\ref{fig:validation-process}, all the output from the simulation and measurement experiments (including the input experiments) are used in the analysis step, the one in which our validation conclusion comes from. Next, both measurement and input experiments are described in detail later in Section~\ref{subsec:measurement-experiments}, and Section~\ref{subsec:simulation-experiments} focused solely in describe the simulation experiment.

\subsection{Measurement Experiments}
\label{subsec:measurement-experiments}

\begin{figure}[ht]
  \centering
  \includegraphics[width=1\linewidth]{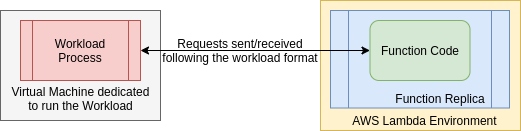}  
\caption{Experiment environment for both measurement experiments. The Workload Generator (WG) running in a different virtual machine sents requests to a function replica in the AWS Lambda environment and collect the service time (time took to process the function code). The WG sends the requests following a predefined distribution (sequential or Poisson, depending on the type of the measurement experiment).}
\label{fig:measurements-environment}
\end{figure}

As introduced in Section ~\ref{sec:methodology}, our study contains two types of measurement experiments, the input experiments and the measurement experiments. The input experiments are the measurement experiments executed aiming to generate inputs to the simulation and the measurement experiments are the ones executed in order to generate results for comparison purposes. The input experiments focused on generating the simulator inputs and are essential to the understanding of the function behavior in the AWS Lambda environment. It also was used to generate the workload used in the measurement experiments, since the workload settings required the mean of the function response time. These experiments were carefully design to mitigate interference from external sources to focus only on the runtime environment. Thus, we argue that the observed response time is mainly due to the function service time, i.e., the function code plus runtime procedures, and the cold start. For both two measurement experiments, the illustration in Figure~\ref{fig:measurements-environment} emphasizes the environment of these measurements. The following sections describes in details the measurement experiments executed for simulation input and validation comparison purposes. 

\subsubsection{Simulation Inputs}
\label{subsec:simulation-inputs}

In order to collect the simulation inputs, our experiments measured the performance of a resize function function~\cite{github:gci-faas-sim:handler}, which is a popular application used in performance evaluation\cite{oakes2018sock, akkus2018sand}. Each function invocation reduces a 560KB sized image to $10\%$ of its original size. To reduce the effect of I/O transfers and network fluctuations in experiment results, the original image is loaded and stored in memory during the function startup, and the resized images are not sent back to the client, only the service time (time to execute the function code) to resize the image. Each measurement experiment observes the behavior of a single instance of the resize function application running on the AWS Lambda functions environment.

For each run, a process running in a different virtual machine sends a sequence (which means that a request was sent only after the receiving of the response of the previous one) of $5.000$ requests, stores its response times and status codes. Our analysis ignored the response time of the $250$ ($5\%$) first requests of each run to soften the effects of the warmup~\cite{Blackburn:2008:WUS:1378704.1378723}. We executed a total of $32$ runs. And between each run, we waited one hour to make sure a new instance is created and the effects of cold start properly accounted for each measurement experiment.

The measurement experiments for simulation inputs were also useful to define the Poisson distribution used in the~\ref{subsec:measurements-for-validation}. Thus, we used the mean of the response time of the function invocations measured for the simulation inputs as the lambda value of the Poisson distribution used in the measurements experiments for validation purposes. This way, the mean of the inter-arrival of function invocations is set to be equal to the mean of the response time of the function. Additionally, these measurement experiments were also used for sanity and any additional analysis that require further understanding of the response time of the function behavior in the AWS Lambda environment.

\subsubsection{Measurements for Validation}
\label{subsec:measurements-for-validation}

The measurement experiments for comparison and validation purposes consist of an execution of a workload in a different virtual machine using a Poisson distribution to define the time between each function invocation. Differently from the measurements in~\ref{subsec:simulation-inputs}, in this workload we surely have concurrency occurring and triggering new function instances, having the concurrency level depending on the distribution intensity. As well as in~\ref{subsec:simulation-inputs}, this measurement experiment also used the resizer function~\cite{github:gci-faas-sim:handler}, in which each function invocation resizes a 560KB image to $10\%$ of its original size, and all procedures to avoid I/O transfers and network fluctuations in the experiments was also used here.

For each run, a workload sent $20.000$ requests to the function replica in AWS lambda environment with the inter-arrival following the Poisson distribution, then stores the responses and status codes. In order to avoid warmup~\cite{Blackburn:2008:WUS:1378704.1378723} influence, we ignored the response time of the $1.000$ ($5\%$) first requests of each run. Still, as well as in~\ref{subsec:simulation-inputs}, between each run we waited for enough time (more than one hour) to make sure that the function replica was not ready to run at the beginning of each experiment. A total of 8 simulations, 4 for each lambda value, was executed.

The Poisson distribution used, as pointed in~\ref{subsec:simulation-inputs}, was defined with the lambda value equal to the mean of the response time of the function invocations measured during the measurement experiments for the simulation inputs. By defining the lambda value to that, we ensure that the Poisson distribution periodically generates a function invocation before the previous request is finished, and therefore, the system have to trigger a new function replica, increasing the concurrency level.

\subsection{Simulation experiments}
\label{subsec:simulation-experiments}
In order to validate the simulation model used in the previous work~\cite{quaresma-sbacpad} and described in Section~\ref{subsec:simulation-model}, we executed a series of simulation experiments. The main objective of this experiment is to simulate the measurement experiments described in Section~\ref{subsec:measurements-for-validation} to later compare the results and validate the simulator. This experiment simulated the sending of $20.000$ function invocations following the Poisson distribution, using the same settings of the respective measurement experiments. In order to avoid warmup~\cite{Blackburn:2008:WUS:1378704.1378723} influence, for each run, we removed the first $1.000$ ($5\%$) service times simulated. A total of 4 simulations run was executed.

Regarding the input of the simulator, to each possible function instance the simulation might have simulate, a respective input file previously measured is required. These input files were generated through the measurement experiments for the simulation inputs, as described in~\ref{subsec:simulation-inputs}. A total of $32$ input files was used in all simulation experiments to be reproduced among all function replicas created during simulation. Each of this files contained a total of $5.000$ tuples as described in Section~\ref{sec:instance}. Among the several outputs generated from the simulator, in our analysis, we focused mainly on the service time to the validation, but we also used instances and cold start statistics for comprehension concerns.

There are two limitations in our model. One is that the simulation might trigger more function instances than the input files passed, and the other is that the file might not have enough entries for the whole simulation. In the first case, if a new function instance is created and there is no unused input file, the simulator will reuse the one that was used less recently. In the latter case, if the number of entries within the input file is not enough for the whole simulation, the simulation will reset the file iteration to just after the cold start entry (to avoid multiples cold starts in the same function instance).

\section{Results}
\label{sec:results}

In this section, we describe the results attained in the validation process described in section~~\ref{subsec:validation-process}. Briefly, by comparing the results of the measurement experiments with the ones from the simulation experiments, we observed that the simulated model was effective in simulating the shape of the response time distribution of the AWS Lambda environment. We also noticed an increase in the response time mean, median, and percentiles. However, this positive shift of value in the response time does not impact the validity of the conclusions in the study executed in the previous work~\cite{quaresma-sbacpad} since the study relied only on the distribution shape for the conclusions, as talked in Section~\ref{subsec:previous-work}.

The remaining of this section presents and discuss the results attained in detail. We start our analysis by evaluating the Empirical Cumulative Distribution Function (ECDF) of the simulation experiments and the measurements experiments for input and validation purposes. Then, we studied the shape of the response time distribution generated in simulation and measurement experiments results.Finally, we compared statistics such as mean, median, and percentiles for all experiments, and shared the sanity checks made in order to testify the results observed. 

\begin{figure}[ht]
  \centering
  \includegraphics[width=1\linewidth]{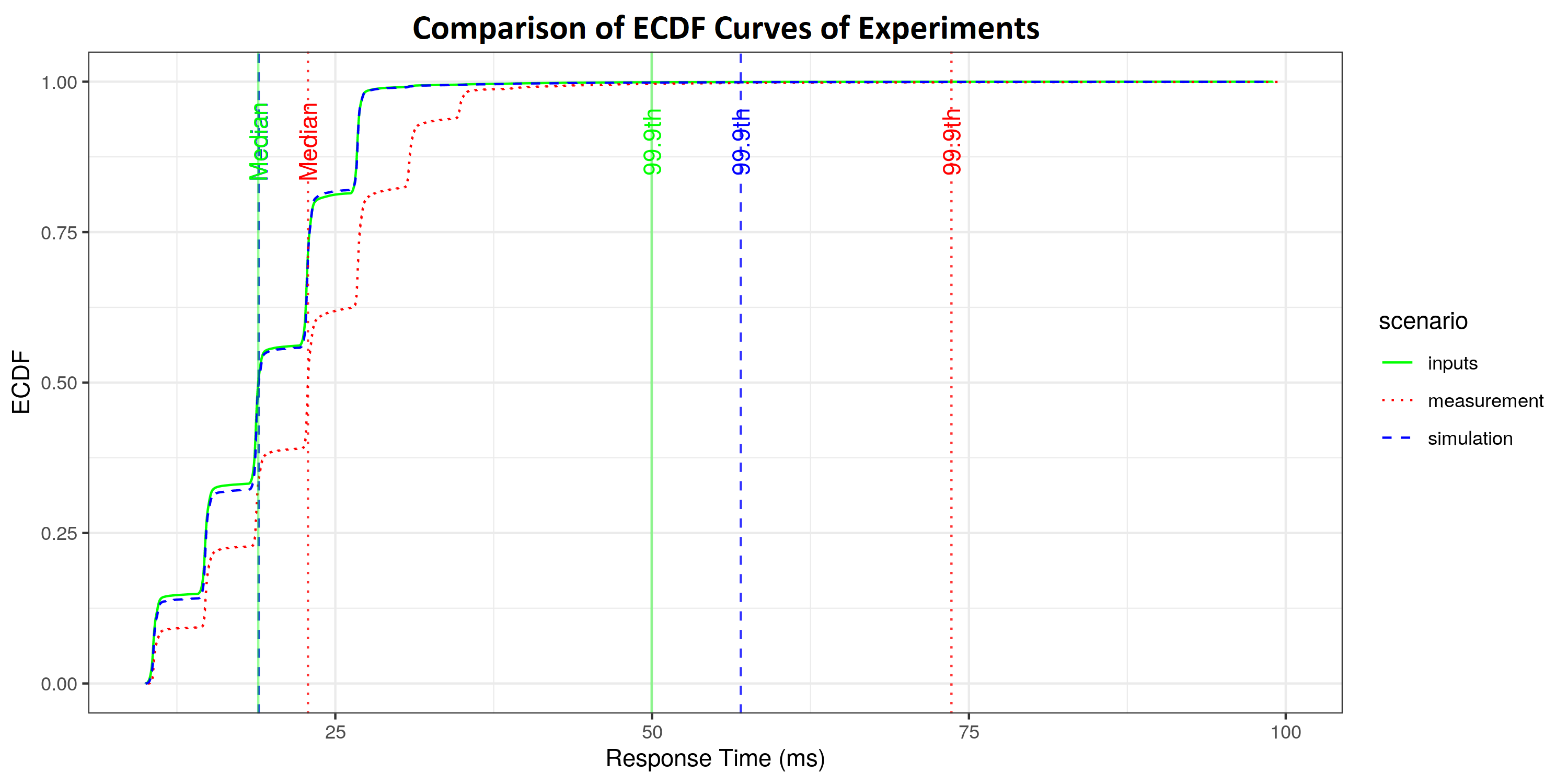}  
\caption{The Empirical Cumulative Distribution Function (ECDF) curves of all the simulation and measurement experiments. While the ECDF of the simulation and input measurements experiments are quite identical, the measurement for comparison purposes experiments shares the same shape but has higher values of response time.}
\label{fig:ecdf}
\end{figure}

Figure~\ref{fig:ecdf} presents the ECDF of the response time for the results attained in the simulation and measurements experiments. The vertical lines in the image indicate the median and percentile $99.9th$ of each distribution of each experiment results. Solid green line represents the results from input experiments, blue dashed lines represents the results from simulation experiments and red dotted lines means results from the measurement experiments. As we can see, in Figure~\ref{fig:ecdf} while simulation and input experiments share similar distribution shape and values, being likely identical curves, the measurement ECDF curve has higher response time values, even with a similar distribution shape. In the median, we see an overlap between the input and simulation experiments, indicating that both have the same median. Differently, in the percentile $99.9th$, we have a minimal difference between the simulation and input experiments, with the simulation response time being slightly higher over $5$ms. Regarding the median and percentiles of the measurement experiments, both the median of the response time is bit higher if compared to the ones obtained in the simulation and inputs experiments results. The measurement response time median if slightly greater (less than $5$ms) than the ones from simulation and inputs, while the percentile $99.9th$ is considerably higher (more than $20$ms).

\begin{figure}[ht]
  \centering
  \includegraphics[width=1\linewidth]{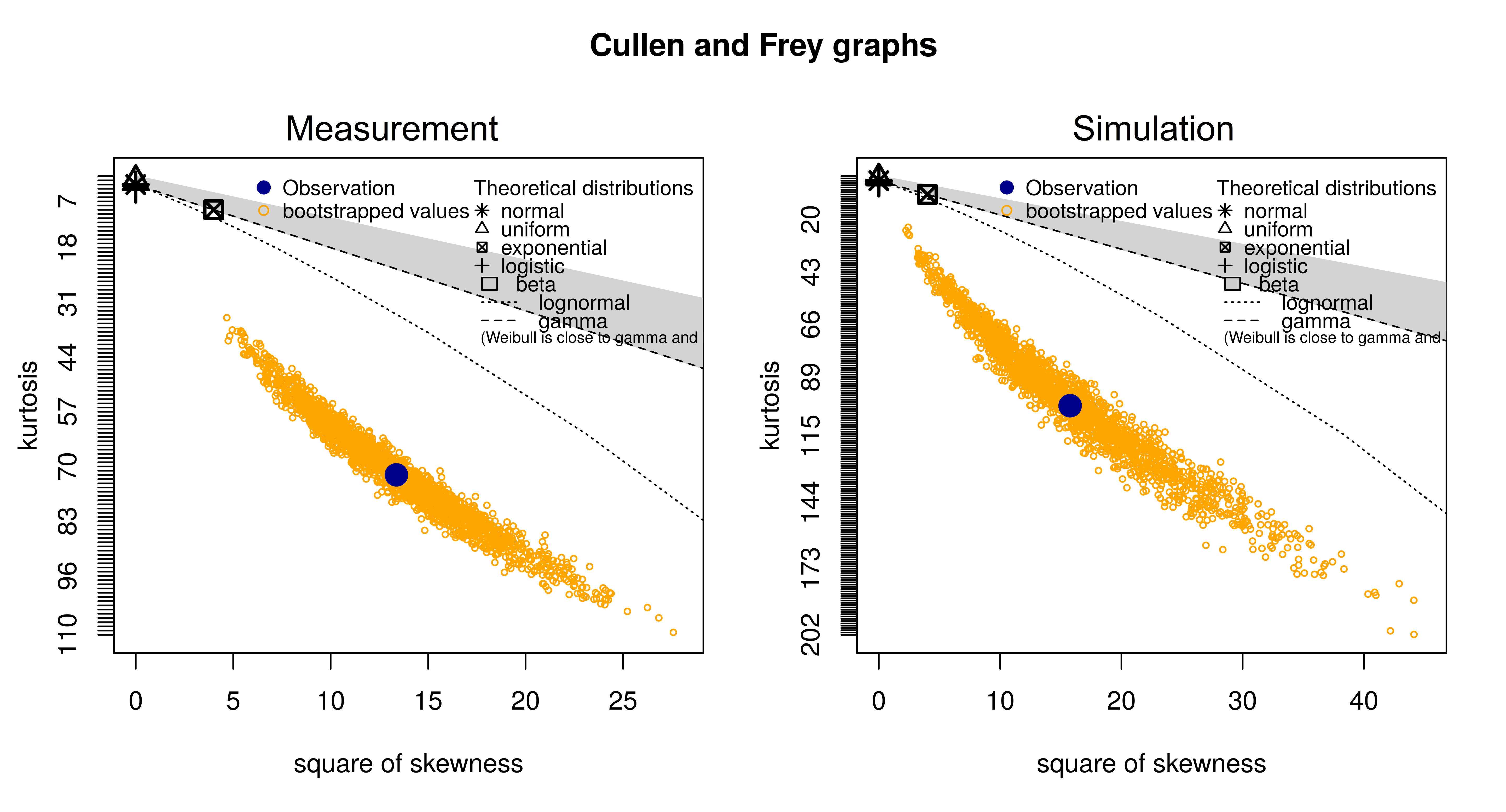}  
\caption{Cullen and Frey graphs for both simulation and measurement experiments results. The skewness of the distribution indicates how symmetrical the distribution is, being skewness of $0$ a perfect symmetrical distribution~\cite{SkewnessandKurtosis}. Kurtosis stands for a measure of whether the distribution is peaked, i.e., the distribution has most of the values in the center)~\cite{PLS_SEM-skewnessandkurtosis}. In the graph, both simulation and measurement experiments shares similarly values for Skewness and Kurtosis.}
\label{fig:cullen-frey-graph}
\end{figure}

In Figure~\ref{fig:cullen-frey-graph}, we have a Cullen and Grey Graph, specifically the skewness-kurtosis plot, For the simulation and measurement experiments. While Skewness is basically a measure of the relative size of the two tails of the distribution, the Kurtosis is the measure of the combined sizes of both tails~\cite{SkewnessandKurtosis}. Regarding skewness, as symmetric the distribution tails are, the closer to $0$ the Skewness is. On the other side, the kurtosis measure is a measures of how peaked a distribution is. It is often measured comparing to the Kurtosis of the normal distribution, which is $3$. So, as closer to $3$ the Kurtosis of a distribution is, the closer it is to a normal distribution. If you need to evaluate a normal distribution under the skewness-kurtosis graphic, is expected skewness values closer to $0$ and kurtosis values closer to $3$.

As we can see in the Figure~\ref{fig:cullen-frey-graph}, both simulation and measurement experiments have very similar values for skewness and kurtosis, which indicates that the distribution of response time of both experiments is the same. Both distributions are highly tailered distributions, with a fairly large range in peak variation. We ensured that the workload applied in both scenarios was the same by studying how the workload concurrency triggered function instances in both experiments. The triggering of function instances might heavily impact the distribution shape due to cold starts, so peak of high concurrency level along the workload would trigger more cold start.

We concluded, by analyzing the simulator and CloudWatch~\cite{aws-cloudwatch-page} logs that both experiment had the same peaks of concurrency and all cold start happened in the begging of the benchmanking. Since this studied focused in validade whether the simulation model is valid for the designed scarios of the previous work~\cite{quaresma-sbacpad}, we argue that this does not impact the validity of the conclusion of the previous study. However, we assert that for a more generalist usage of this simulation model, a more realistic workload would be required.

\begin{table}[htbp]
\caption{Presents a summary table of the percentiles under the confidence interval of $95\%$ calculated for the simulation and measurement experiments. The table shows that the response time of the distributions studied is statistically different and that the measurement experiments has higher response time values.}
\label{tab:percentiles}
\begin{center}
\setlength\tabcolsep{4pt}
\begin{tabular}{lll}
\centering
Percentile. & Measurement (ms) & Simulation (ms)\\
$50^{th}$ &[$22.83$, $22.84$]& [$18.93$, $18.97$]\\
$95^{th}$ & [$34.76$, $34.79$]& [$26.89$, $26.91$]\\
$99^{th}$ & [$39.12$, $39.84$]& [$29.03$, $30.68$]\\
$99.9^{th}$ & [$69.14$, $79.70$]& [$53.29$, $60.28$]\\
\end{tabular}
\end{center}
\end{table}

Table~\ref{tab:percentiles} shows the percentiles values for the measurement and simulation experiments, ranging from $50th$ to $99.9th$. First, we noticed that for all percentiles and for the confidence interval of $95\%$, all percentiles studied for simulation and measurement experiments are different. Also, the higher the percentile, the higher is the response time of the measurement experiment, just as indicated in the Figure~\ref{fig:ecdf} in the median and percentile indicators. The mean of the response time of the simulation input is the same as the ones from input experiments, around $19$ms, while the mean of response time from measurement experiments is about $22$ms. According to the range of the confidence interval, the difference between the mean of the simulation and measurement experiments goes from $3.86$ms to $3.91$ms. Respecting to the percentile of $99.9th$, the difference between the simulation and measurement experiments varies from $8.86$ms to $26.41$ms. We argue that these differences in low enough to be ignored since it these small variations do not critically change the distribution shape as shown in Figure~\ref{fig:ecdf} and explained in Figure~\ref{fig:cullen-frey-graph}.

To the understanding of the results attained, a fundamental process in the analysis phase is the sanity check. In this process, we executed additional test cases, experimental scenarios, and studied logs from the AWS CloudWatch. We did so in order to understand why the increase of concurrency was also increasing the service time of function execution in the AWS Lambda environment. We noticed that in our results following, as higher our concurrency level was, the higher was the impact in the service time, but not proportionally. A workload with doubled concurrency level implied only plus $3$ to $4$ milliseconds in the mean of service time. We checked this statistics in the CloudWatch dashboard, and the CloudWatch logs indicated the same as ours. This was consistent and happened across all our experiment repetitions. We argue that the workload increased concurrency probably caused an overhead in the AWS Environment and slighly impacted the function execution, even running in isolated containers, due to the multi-tenancy characteristic of the AWS Lambda platform~\cite{multi-tenancy}.
\section{Conclusion}
\label{sec:conclusion}

In the previous work~\cite{quaresma-sbacpad}, as introduced in Section~\ref{subsec:simulation-model}, during the process of modeling the simulator, we defined an assumption in order to simplify the simulation model. We assumed that the input files used in the simulator would have a similar response time distribution shape. The discussion of the Figure~\ref{fig:ecdf} and Table~\ref{tab:percentiles} showed that the measurement experiments presented slightly higher values for some percentiles, mean, and median. This difference in the response time measured might be justified by the overhead introduced in the provider environment due to the increased concurrency generated by the workload. Still, as shown in the discussion related to Figure~\ref{fig:cullen-frey-graph}, the simulation model is highly powerful in simulating the response time distribution shape, proving the assumption used, which was essential to the scope of the previous work~\cite{quaresma-sbacpad}.

By using the predictive validation technique, we concluded that the simulation model used in the previous work~\cite{quaresma-sbacpad}, and the conclusions from this study, was valid for the scope of the work. Although we noticed differences in the response time values of percentiles, mean and median, these differences were not impactful enough to invalidate the assumption used and, therefore, can not invalidate the simulation model for the scoped it was used. Furthermore, the validation revealed that the simulation model is highly accurate in simulating the response time distribution shape. However, since this work focused only on the validation of the simulation model used in the Controlling Garbage Collection and Request Admission to Improve Performance of FaaS Applications~\cite{quaresma-sbacpad}, there are still open problems to tackle as future work to further validation of the simulation model on a generalist usage scope.

As listed in the previous work, we argue that for generic use of the simulation model we need to validate the model using a greater variety of representative functions. We also argue that a generic validation of the model requires a greater variety of platform platforms as reference, including public vendor such as Google Cloud Function~\cite{google-cloud-functions-page}, Microsoft Azure Function~\cite{azure-function-page} and the already used AWS Lambda Function~\cite{aws-function-page}, and also include open-source platforms such as OpenFaaS~\cite{openfaas-page}, Fission~\cite{fission-page}, and Kubeless~\cite{kubeless-page}. Finally, the complete validation of the simulation model requires a realistic workload, which was already provided in the Serverless in the Wild characterization study~\cite{azure-serverless-in-the-wild}.

\begin{acks}
To Thiago Emmanuel and Daniel Fireman, the two who have strongly contributed to my academic education and the development of this work. To Universidade Federal de Campina Grande (UFCG) for providing the foundations and resources required for my education throughout the graduation period. To Laboratório de Sistemas Distribuídos (LSD), the laboratory that gave me the opportunity to work in the academic area and introduce myself in the research field. To VTEX, for providing a laptop and knowledge, through the VTEX Lab program, that allowed me to orchestrate all experiments and execute the analysis upon the results. Lastly, Amazon Web Services (AWS), which, by providing the AWS Free Tier program, makes easier and cost safe the execution of the measurement experiments used in this work. 
\end{acks}

\bibliographystyle{ACM-Reference-Format}
\bibliography{bib}


\begin{thebibliography}{38}


\ifx \showCODEN    \undefined \def \showCODEN     #1{\unskip}     \fi
\ifx \showDOI      \undefined \def \showDOI       #1{#1}\fi
\ifx \showISBNx    \undefined \def \showISBNx     #1{\unskip}     \fi
\ifx \showISBNxiii \undefined \def \showISBNxiii  #1{\unskip}     \fi
\ifx \showISSN     \undefined \def \showISSN      #1{\unskip}     \fi
\ifx \showLCCN     \undefined \def \showLCCN      #1{\unskip}     \fi
\ifx \shownote     \undefined \def \shownote      #1{#1}          \fi
\ifx \showarticletitle \undefined \def \showarticletitle #1{#1}   \fi
\ifx \showURL      \undefined \def \showURL       {\relax}        \fi
\providecommand\bibfield[2]{#2}
\providecommand\bibinfo[2]{#2}
\providecommand\natexlab[1]{#1}
\providecommand\showeprint[2][]{arXiv:#2}

\bibitem[\protect\citeauthoryear{??}{Ter}{1979}]%
        {Terminology-for-model-credibility}
 \bibinfo{year}{1979}\natexlab{}.
\newblock \showarticletitle{Terminology for model credibility}.
\newblock   \bibinfo{volume}{32} (\bibinfo{date}{3} \bibinfo{year}{1979}),
  \bibinfo{pages}{103–104}.
\newblock
\urldef\tempurl%
\url{https://doi.org/10.1177/003754977903200304}
\showDOI{\tempurl}


\bibitem[\protect\citeauthoryear{Akkus, Chen, Rimac, Stein, Satzke, Beck,
  Aditya, and Hilt}{Akkus et~al\mbox{.}}{2018}]%
        {akkus2018sand}
\bibfield{author}{\bibinfo{person}{Istemi~Ekin Akkus},
  \bibinfo{person}{Ruichuan Chen}, \bibinfo{person}{Ivica Rimac},
  \bibinfo{person}{Manuel Stein}, \bibinfo{person}{Klaus Satzke},
  \bibinfo{person}{Andre Beck}, \bibinfo{person}{Paarijaat Aditya}, {and}
  \bibinfo{person}{Volker Hilt}.} \bibinfo{year}{2018}\natexlab{}.
\newblock \showarticletitle{{SAND:} Towards High-Performance Serverless
  Computing}. In \bibinfo{booktitle}{\emph{2018 {USENIX} Annual Technical
  Conference, {USENIX} {ATC} 2018, Boston, MA, USA, July 11-13, 2018}},
  \bibfield{editor}{\bibinfo{person}{Haryadi~S. Gunawi} {and}
  \bibinfo{person}{Benjamin Reed}} (Eds.). \bibinfo{publisher}{{USENIX}
  Association}, \bibinfo{pages}{923--935}.
\newblock
\urldef\tempurl%
\url{https://www.usenix.org/conference/atc18/presentation/akkus}
\showURL{%
\tempurl}


\bibitem[\protect\citeauthoryear{(AWS)}{(AWS)}{[n. d.]a}]%
        {aws-cloudwatch-page}
\bibfield{author}{\bibinfo{person}{Amazon Web~Services (AWS)}.}
  \bibinfo{year}{[n. d.]}\natexlab{a}.
\newblock \bibinfo{title}{Amazon CloudWatch: Observability of your AWS
  resources and applications on AWS and on-premises.}
\newblock \bibinfo{howpublished}{https://aws.amazon.com/cloudwatch/}.
\newblock
\newblock
\shownote{Online; Accessed: 2021-15-03.}


\bibitem[\protect\citeauthoryear{(AWS)}{(AWS)}{[n. d.]b}]%
        {aws-function-page}
\bibfield{author}{\bibinfo{person}{Amazon Web~Services (AWS)}.}
  \bibinfo{year}{[n. d.]}\natexlab{b}.
\newblock \bibinfo{title}{AWS Lambda: Run code without thinking about servers
  or clusters. Only pay for what you use.}
\newblock \bibinfo{howpublished}{https://aws.amazon.com/lambda/}.
\newblock
\newblock
\shownote{Online; Accessed: 2021-15-03.}


\bibitem[\protect\citeauthoryear{(AWS)}{(AWS)}{[n. d.]c}]%
        {aws:lambda-concurrency-level}
\bibfield{author}{\bibinfo{person}{Amazon Web~Services (AWS)}.}
  \bibinfo{year}{[n. d.]}\natexlab{c}.
\newblock \bibinfo{title}{Managing concurrency for a Lambda function}.
\newblock
  \bibinfo{howpublished}{https://docs.aws.amazon.com/lambda/latest/dg/configuration-concurrency.html}.
\newblock
\newblock
\shownote{Online; Accessed: 2021-05-09.}


\bibitem[\protect\citeauthoryear{Azure}{Azure}{[n. d.]}]%
        {azure-function-page}
\bibfield{author}{\bibinfo{person}{Microsoft Azure}.} \bibinfo{year}{[n.
  d.]}\natexlab{}.
\newblock \bibinfo{title}{Azure Functions: More than just event-driven
  serverless compute}.
\newblock
  \bibinfo{howpublished}{https://azure.microsoft.com/en-us/services/functions/}.
\newblock
\newblock
\shownote{Online; Accessed: 2021-15-03.}


\bibitem[\protect\citeauthoryear{Blackburn, McKinley, Garner, Hoffmann, Khan,
  Bentzur, Diwan, Feinberg, Frampton, Guyer, Hirzel, Hosking, Jump, Lee, Moss,
  Phansalkar, Stefanovic, VanDrunen, von Dincklage, and Wiedermann}{Blackburn
  et~al\mbox{.}}{2008}]%
        {Blackburn:2008:WUS:1378704.1378723}
\bibfield{author}{\bibinfo{person}{Stephen~M. Blackburn},
  \bibinfo{person}{Kathryn~S. McKinley}, \bibinfo{person}{Robin Garner},
  \bibinfo{person}{Chris Hoffmann}, \bibinfo{person}{Asjad~M. Khan},
  \bibinfo{person}{Rotem Bentzur}, \bibinfo{person}{Amer Diwan},
  \bibinfo{person}{Daniel Feinberg}, \bibinfo{person}{Daniel Frampton},
  \bibinfo{person}{Samuel~Z. Guyer}, \bibinfo{person}{Martin Hirzel},
  \bibinfo{person}{Antony~L. Hosking}, \bibinfo{person}{Maria Jump},
  \bibinfo{person}{Han Lee}, \bibinfo{person}{J.~Eliot~B. Moss},
  \bibinfo{person}{Aashish Phansalkar}, \bibinfo{person}{Darko Stefanovic},
  \bibinfo{person}{Thomas VanDrunen}, \bibinfo{person}{Daniel von Dincklage},
  {and} \bibinfo{person}{Ben Wiedermann}.} \bibinfo{year}{2008}\natexlab{}.
\newblock \showarticletitle{Wake Up and Smell the Coffee: Evaluation
  Methodology for the 21st Century}.
\newblock \bibinfo{journal}{\emph{Commun. ACM}} \bibinfo{volume}{51},
  \bibinfo{number}{8} (\bibinfo{date}{Aug.} \bibinfo{year}{2008}),
  \bibinfo{pages}{83--89}.
\newblock
\showISSN{0001-0782}
\urldef\tempurl%
\url{https://doi.org/10.1145/1378704.1378723}
\showDOI{\tempurl}


\bibitem[\protect\citeauthoryear{Cloud}{Cloud}{[n. d.]}]%
        {google-cloud-functions-page}
\bibfield{author}{\bibinfo{person}{Google Cloud}.} \bibinfo{year}{[n.
  d.]}\natexlab{}.
\newblock \bibinfo{title}{Cloud Functions}.
\newblock \bibinfo{howpublished}{https://cloud.google.com/functions}.
\newblock
\newblock
\shownote{Online; Accessed: 2021-15-03.}


\bibitem[\protect\citeauthoryear{da~Silva and Carvalho}{da~Silva and
  Carvalho}{2019}]%
        {marcus:sbrc2019:serverless}
\bibfield{author}{\bibinfo{person}{Matheus~Nicolas da Silva} {and}
  \bibinfo{person}{Marcus Carvalho}.} \bibinfo{year}{2019}\natexlab{}.
\newblock \showarticletitle{Análise de Mecanismos de Serverless Computing em
  Ambientes de Nuvens Computacionais}. In \bibinfo{booktitle}{\emph{Anais
  Estendidos do XXXVII Simpósio Brasileiro de Redes de Computadores e Sistemas
  Distribuídos}}. \bibinfo{publisher}{SBC}, \bibinfo{address}{Porto Alegre,
  RS, Brasil}, \bibinfo{pages}{225--232}.
\newblock
\urldef\tempurl%
\url{https://doi.org/10.5753/sbrc_estendido.2019.7791}
\showDOI{\tempurl}


\bibitem[\protect\citeauthoryear{Eyk, Iosup, Grohmann, Eismann, Bauer,
  Versluis, Toader, Schmitt, Herbst, and Abad}{Eyk et~al\mbox{.}}{2019}]%
        {SPEC-RG:reference-for-faas}
\bibfield{author}{\bibinfo{person}{Erwin Eyk}, \bibinfo{person}{Alexandru
  Iosup}, \bibinfo{person}{Johannes Grohmann}, \bibinfo{person}{Simon Eismann},
  \bibinfo{person}{André Bauer}, \bibinfo{person}{Laurens Versluis},
  \bibinfo{person}{Lucian Toader}, \bibinfo{person}{Norbert Schmitt},
  \bibinfo{person}{Nikolas Herbst}, {and} \bibinfo{person}{Cristina Abad}.}
  \bibinfo{year}{2019}\natexlab{}.
\newblock \showarticletitle{The SPEC-RG Reference Architecture for FaaS: From
  Microservices and Containers to Serverless Platforms}.
\newblock \bibinfo{journal}{\emph{IEEE Internet Computing}}
  \bibinfo{volume}{PP} (\bibinfo{date}{11} \bibinfo{year}{2019}),
  \bibinfo{pages}{1--1}.
\newblock
\urldef\tempurl%
\url{https://doi.org/10.1109/MIC.2019.2952061}
\showDOI{\tempurl}


\bibitem[\protect\citeauthoryear{Farrington, Nembhard, Sturrock, Evans, and
  Sargent}{Farrington et~al\mbox{.}}{2000}]%
        {Validation-And-Verification-Of-Simulation-Models}
\bibfield{author}{\bibinfo{person}{Phillip Farrington},
  \bibinfo{person}{Harriet Nembhard}, \bibinfo{person}{D. Sturrock},
  \bibinfo{person}{G. Evans}, {and} \bibinfo{person}{Robert Sargent}.}
  \bibinfo{year}{2000}\natexlab{}.
\newblock \showarticletitle{Validation And Verification Of Simulation Models}.
\newblock  (\bibinfo{date}{02} \bibinfo{year}{2000}).
\newblock


\bibitem[\protect\citeauthoryear{{Fireman}, {Brunet}, {Lopes}, {Quaresma}, and
  {Pereira}}{{Fireman} et~al\mbox{.}}{2018}]%
        {fireman-cloudcom}
\bibfield{author}{\bibinfo{person}{D. {Fireman}}, \bibinfo{person}{J.
  {Brunet}}, \bibinfo{person}{R. {Lopes}}, \bibinfo{person}{D. {Quaresma}},
  {and} \bibinfo{person}{T.~E. {Pereira}}.} \bibinfo{year}{2018}\natexlab{}.
\newblock \showarticletitle{Improving Tail Latency of Stateful Cloud Services
  via GC Control and Load Shedding}. In \bibinfo{booktitle}{\emph{2018 IEEE
  International Conference on Cloud Computing Technology and Science
  (CloudCom)}}. \bibinfo{publisher}{{IEEE} Computer Society},
  \bibinfo{pages}{121--128}.
\newblock
\showISSN{2330-2186}
\urldef\tempurl%
\url{https://doi.org/10.1109/CloudCom2018.2018.00034}
\showDOI{\tempurl}


\bibitem[\protect\citeauthoryear{Fireman, Lopes, and Brunet~Monteiro}{Fireman
  et~al\mbox{.}}{2017}]%
        {FiremanSBRC2017}
\bibfield{author}{\bibinfo{person}{Daniel Fireman}, \bibinfo{person}{Raquel
  Lopes}, {and} \bibinfo{person}{Jo{\~a}o~Arthur Brunet~Monteiro}.}
  \bibinfo{year}{2017}\natexlab{}.
\newblock \showarticletitle{Using Load Shedding to Fight Tail-Latency on
  Runtime-Based Services}. In \bibinfo{booktitle}{\emph{Brazilian Symposium on
  Computer Networks and Distributed Systems}}.
\newblock


\bibitem[\protect\citeauthoryear{Fireman and Quaresma}{Fireman and
  Quaresma}{2020a}]%
        {github:gci-faas-sim:handler}
\bibfield{author}{\bibinfo{person}{Daniel Fireman} {and} \bibinfo{person}{David
  Quaresma}.} \bibinfo{year}{2020}\natexlab{a}.
\newblock \bibinfo{title}{GCI FaaS Sim, Handler}.
\newblock
  \bibinfo{howpublished}{\url{https://github.com/dfquaresma/gci-faas-sim/blob/master/runtime/thumb-func/src/main/java/com/openfaas/function/Handler.java}}.
\newblock


\bibitem[\protect\citeauthoryear{Fireman and Quaresma}{Fireman and
  Quaresma}{2020b}]%
        {github:gci-simulator}
\bibfield{author}{\bibinfo{person}{Daniel Fireman} {and} \bibinfo{person}{David
  Quaresma}.} \bibinfo{year}{2020}\natexlab{b}.
\newblock \bibinfo{title}{GCI Simulator}.
\newblock
  \bibinfo{howpublished}{\url{https://github.com/gcinterceptor/gci-simulator/tree/master/serverlessgo}}.
\newblock


\bibitem[\protect\citeauthoryear{Fission}{Fission}{[n. d.]}]%
        {fission-page}
\bibfield{author}{\bibinfo{person}{Fission}.} \bibinfo{year}{[n.
  d.]}\natexlab{}.
\newblock \bibinfo{title}{Fission: Open source, Kubernetes-native Serverless
  Framework}.
\newblock \bibinfo{howpublished}{https://fission.io/}.
\newblock
\newblock
\shownote{Online; Accessed: 2021-15-03.}


\bibitem[\protect\citeauthoryear{Guru}{Guru}{[n. d.]}]%
        {A-Cloud-Guru-about-idleness}
\bibfield{author}{\bibinfo{person}{A~Cloud Guru}.} \bibinfo{year}{[n.
  d.]}\natexlab{}.
\newblock \bibinfo{title}{How long does AWS Lambda keep your idle functions
  around before a cold start?}
\newblock
  \bibinfo{howpublished}{https://acloudguru.com/blog/engineering/how-long-does-aws-lambda-keep-your-idle-functions-around-before-a-cold-start}.
\newblock
\newblock
\shownote{Online; Accessed: 2021-05-09.}


\bibitem[\protect\citeauthoryear{Hair, Hult, Ringle, and Sarstedt}{Hair
  et~al\mbox{.}}{2016}]%
        {PLS_SEM-skewnessandkurtosis}
\bibfield{author}{\bibinfo{person}{Joe Hair}, \bibinfo{person}{G.~Tomas~M.
  Hult}, \bibinfo{person}{Christian Ringle}, {and} \bibinfo{person}{Marko
  Sarstedt}.} \bibinfo{year}{2016}\natexlab{}.
\newblock \bibinfo{booktitle}{\emph{A Primer on Partial Least Squares
  Structural Equation Modeling (PLS-SEM), 2nd edition}}.
\newblock
\showISBNx{9781483377445}


\bibitem[\protect\citeauthoryear{Hu}{Hu}{[n. d.]}]%
        {infoq-serverless-popularity}
\bibfield{author}{\bibinfo{person}{Vivian Hu}.} \bibinfo{year}{[n.
  d.]}\natexlab{}.
\newblock \bibinfo{title}{Serverless Days 2020 Looks at Future of Serverless
  Architecture}.
\newblock
  \bibinfo{howpublished}{https://www.infoq.com/news/2020/07/future-serverless-architecture/}.
\newblock
\newblock
\shownote{Online; Accessed: 2021-05-12.}


\bibitem[\protect\citeauthoryear{{Kritikos} and {Skrzypek}}{{Kritikos} and
  {Skrzypek}}{2018}]%
        {A-Review-of-Serverless-Frameworks}
\bibfield{author}{\bibinfo{person}{K. {Kritikos}} {and} \bibinfo{person}{P.
  {Skrzypek}}.} \bibinfo{year}{2018}\natexlab{}.
\newblock \showarticletitle{A Review of Serverless Frameworks}. In
  \bibinfo{booktitle}{\emph{2018 IEEE/ACM International Conference on Utility
  and Cloud Computing Companion (UCC Companion)}}. \bibinfo{pages}{161--168}.
\newblock


\bibitem[\protect\citeauthoryear{Kubeless}{Kubeless}{[n. d.]}]%
        {kubeless-page}
\bibfield{author}{\bibinfo{person}{Kubeless}.} \bibinfo{year}{[n.
  d.]}\natexlab{}.
\newblock \bibinfo{title}{Kubeless: The Kubernetes Native Serverless
  Framework}.
\newblock \bibinfo{howpublished}{https://kubeless.io/}.
\newblock
\newblock
\shownote{Online; Accessed: 2021-15-03.}


\bibitem[\protect\citeauthoryear{Kumar}{Kumar}{2019}]%
        {Kumar2019}
\bibfield{author}{\bibinfo{person}{Manoj Kumar}.}
  \bibinfo{year}{2019}\natexlab{}.
\newblock \showarticletitle{Serverless Architectures Review, Future Trend and
  the Solutions to Open Problems}.
\newblock   \bibinfo{volume}{6} (\bibinfo{date}{03} \bibinfo{year}{2019}),
  \bibinfo{pages}{8}.
\newblock
\urldef\tempurl%
\url{https://doi.org/10.12691/ajse-6-1-1}
\showDOI{\tempurl}


\bibitem[\protect\citeauthoryear{Lee, Satyam, and Fox}{Lee
  et~al\mbox{.}}{2018}]%
        {Evaluation-of-Production-Serverless-Computing-Environments}
\bibfield{author}{\bibinfo{person}{Hyungro Lee}, \bibinfo{person}{Kumar
  Satyam}, {and} \bibinfo{person}{Geoffrey Fox}.}
  \bibinfo{year}{2018}\natexlab{}.
\newblock \showarticletitle{Evaluation of Production Serverless Computing
  Environments}. \bibinfo{pages}{442--450}.
\newblock
\urldef\tempurl%
\url{https://doi.org/10.1109/CLOUD.2018.00062}
\showDOI{\tempurl}


\bibitem[\protect\citeauthoryear{Manner, Endreß, Heckel, and Wirtz}{Manner
  et~al\mbox{.}}{2018}]%
        {coldstartref}
\bibfield{author}{\bibinfo{person}{Johannes Manner}, \bibinfo{person}{Martin
  Endreß}, \bibinfo{person}{Tobias Heckel}, {and} \bibinfo{person}{Guido
  Wirtz}.} \bibinfo{year}{2018}\natexlab{}.
\newblock \showarticletitle{Cold Start Influencing Factors in Function as a
  Service}.
\newblock
\urldef\tempurl%
\url{https://doi.org/10.1109/UCC-Companion.2018.00054}
\showDOI{\tempurl}


\bibitem[\protect\citeauthoryear{McNeese}{McNeese}{[n. d.]}]%
        {SkewnessandKurtosis}
\bibfield{author}{\bibinfo{person}{Dr.~Bill McNeese}.} \bibinfo{year}{[n.
  d.]}\natexlab{}.
\newblock \bibinfo{title}{Are the Skewness and Kurtosis Useful Statistics?}
\newblock
  \bibinfo{howpublished}{https://www.spcforexcel.com/knowledge/basic-statistics/are-skewness-and-kurtosis-useful-statistics}.
\newblock
\newblock
\shownote{Online; Accessed: 2021-05-12.}


\bibitem[\protect\citeauthoryear{MITRE}{MITRE}{2020}]%
        {mitre-validation}
\bibfield{author}{\bibinfo{person}{MITRE}.} \bibinfo{year}{2020}\natexlab{}.
\newblock \bibinfo{title}{Verification and Validation of Simulation Models}.
\newblock
\newblock
\urldef\tempurl%
\url{https://www.mitre.org/publications/systems-engineering-guide/se-lifecycle-building-blocks/other-se-lifecycle-building-blocks-articles/verification-and-validation-of-simulation-models}
\showURL{%
\tempurl}


\bibitem[\protect\citeauthoryear{Oakes, Yang, Zhou, Houck, Harter,
  Arpaci{-}Dusseau, and Arpaci{-}Dusseau}{Oakes et~al\mbox{.}}{2018}]%
        {oakes2018sock}
\bibfield{author}{\bibinfo{person}{Edward Oakes}, \bibinfo{person}{Leon Yang},
  \bibinfo{person}{Dennis Zhou}, \bibinfo{person}{Kevin Houck},
  \bibinfo{person}{Tyler Harter}, \bibinfo{person}{Andrea~C. Arpaci{-}Dusseau},
  {and} \bibinfo{person}{Remzi~H. Arpaci{-}Dusseau}.}
  \bibinfo{year}{2018}\natexlab{}.
\newblock \showarticletitle{{SOCK:} Rapid Task Provisioning with
  Serverless-Optimized Containers}. In \bibinfo{booktitle}{\emph{2018 {USENIX}
  Annual Technical Conference, {USENIX} {ATC} 2018, Boston, MA, USA, July
  11-13, 2018}}, \bibfield{editor}{\bibinfo{person}{Haryadi~S. Gunawi} {and}
  \bibinfo{person}{Benjamin Reed}} (Eds.). \bibinfo{publisher}{{USENIX}
  Association}, \bibinfo{pages}{57--70}.
\newblock
\urldef\tempurl%
\url{https://www.usenix.org/conference/atc18/presentation/oakes}
\showURL{%
\tempurl}


\bibitem[\protect\citeauthoryear{OpenFaaS}{OpenFaaS}{[n. d.]}]%
        {openfaas-page}
\bibfield{author}{\bibinfo{person}{OpenFaaS}.} \bibinfo{year}{[n.
  d.]}\natexlab{}.
\newblock \bibinfo{title}{OpenFaaS: Serverless Functions, Made Simple}.
\newblock \bibinfo{howpublished}{https://www.openfaas.com/}.
\newblock
\newblock
\shownote{Online; Accessed: 2021-15-03.}


\bibitem[\protect\citeauthoryear{Palade, Kazmi, and Clarke}{Palade
  et~al\mbox{.}}{2019}]%
  {An-Evaluation-of-Open-Source-Serverless-Computing-Frameworks-Support-at-the-Edge}
\bibfield{author}{\bibinfo{person}{Andrei Palade}, \bibinfo{person}{Aqeel
  Kazmi}, {and} \bibinfo{person}{Siobhán Clarke}.}
  \bibinfo{year}{2019}\natexlab{}.
\newblock \showarticletitle{An Evaluation of Open Source Serverless Computing
  Frameworks Support at the Edge}.
\newblock
\urldef\tempurl%
\url{https://doi.org/10.1109/SERVICES.2019.00057}
\showDOI{\tempurl}


\bibitem[\protect\citeauthoryear{Quaresma, Fireman, and Pereira}{Quaresma
  et~al\mbox{.}}{2020}]%
        {quaresma-sbacpad}
\bibfield{author}{\bibinfo{person}{David Quaresma}, \bibinfo{person}{Daniel
  Fireman}, {and} \bibinfo{person}{Thiago~Emmanuel Pereira}.}
  \bibinfo{year}{2020}\natexlab{}.
\newblock \showarticletitle{Controlling Garbage Collection and Request
  Admission to Improve Performance of FaaS Applications}. In
  \bibinfo{booktitle}{\emph{2020 IEEE 32nd International Symposium on Computer
  Architecture and High Performance Computing (SBAC-PAD)}}.
  \bibinfo{pages}{175--182}.
\newblock
\urldef\tempurl%
\url{https://doi.org/10.1109/SBAC-PAD49847.2020.00033}
\showDOI{\tempurl}


\bibitem[\protect\citeauthoryear{Rama}{Rama}{[n. d.]}]%
        {awsinsider-was-lambda-popularity}
\bibfield{author}{\bibinfo{person}{Gladys Rama}.} \bibinfo{year}{[n.
  d.]}\natexlab{}.
\newblock \bibinfo{title}{Report: AWS Lambda Popular Among Enterprises,
  Container Users}.
\newblock
  \bibinfo{howpublished}{https://awsinsider.net/articles/2020/02/04/aws-lambda-usage-profile.aspx}.
\newblock
\newblock
\shownote{Online; Accessed: 2021-05-12.}


\bibitem[\protect\citeauthoryear{RedHat}{RedHat}{[n. d.]}]%
        {redhat:what-is-faas}
\bibfield{author}{\bibinfo{person}{RedHat}.} \bibinfo{year}{[n.
  d.]}\natexlab{}.
\newblock \bibinfo{title}{Cloud-Native Applications:What Is
  Function-As-A-Service (Faas)?}
\newblock
  \bibinfo{howpublished}{https://www.redhat.com/en/topics/cloud-native-apps/what-is-faas}.
\newblock
\newblock
\shownote{Online; Accessed: 2021-05-07.}


\bibitem[\protect\citeauthoryear{Robinson}{Robinson}{2000}]%
        {about-validation-and-simulation-concepts}
\bibfield{author}{\bibinfo{person}{Stewart Robinson}.}
  \bibinfo{year}{2000}\natexlab{}.
\newblock \showarticletitle{Simulation Model Verification And Validation:
  Increasing The Users' Confidence}.
\newblock \bibinfo{journal}{\emph{Winter Simulation Conference Proceedings}}
  (\bibinfo{date}{05} \bibinfo{year}{2000}).
\newblock
\urldef\tempurl%
\url{https://doi.org/10.1145/268437.268448}
\showDOI{\tempurl}


\bibitem[\protect\citeauthoryear{Sargent}{Sargent}{2009}]%
        {validation-of-simulation-models}
\bibfield{author}{\bibinfo{person}{Robert Sargent}.}
  \bibinfo{year}{2009}\natexlab{}.
\newblock \showarticletitle{Verification and validation of simulation models}.
\newblock \bibinfo{journal}{\emph{Journal of Simulation}}  \bibinfo{volume}{7},
  \bibinfo{pages}{162--176}.
\newblock
\showISBNx{978-1-4244-5770-0}
\urldef\tempurl%
\url{https://doi.org/10.1109/WSC.2009.5429327}
\showDOI{\tempurl}


\bibitem[\protect\citeauthoryear{Science}{Science}{2020}]%
        {multi-tenancy}
\bibfield{author}{\bibinfo{person}{Narrative Science}.}
  \bibinfo{year}{2020}\natexlab{}.
\newblock \bibinfo{title}{How AWS Lambda Changed the Game of Multi-tenancy}.
\newblock
\newblock
\urldef\tempurl%
\url{https://narrativescience.com/resource/blog/how-aws-lambda-changed-the-game-of-multi-tenancy/}
\showURL{%
\tempurl}


\bibitem[\protect\citeauthoryear{Shahrad, Fonseca, Goiri, Chaudhry, Batum,
  Cooke, Laureano, Tresness, Russinovich, and Bianchini}{Shahrad
  et~al\mbox{.}}{2020}]%
        {azure-serverless-in-the-wild}
\bibfield{author}{\bibinfo{person}{Mohammad Shahrad}, \bibinfo{person}{Rodrigo
  Fonseca}, \bibinfo{person}{Íñigo Goiri}, \bibinfo{person}{Gohar Chaudhry},
  \bibinfo{person}{Paul Batum}, \bibinfo{person}{Jason Cooke},
  \bibinfo{person}{Eduardo Laureano}, \bibinfo{person}{Colby Tresness},
  \bibinfo{person}{Mark Russinovich}, {and} \bibinfo{person}{Ricardo
  Bianchini}.} \bibinfo{year}{2020}\natexlab{}.
\newblock \bibinfo{title}{Serverless in the Wild: Characterizing and Optimizing
  the Serverless Workload at a Large Cloud Provider}.
\newblock
\newblock


\bibitem[\protect\citeauthoryear{Silva, Fireman, and Silva}{Silva
  et~al\mbox{.}}{2020}]%
        {feitosa-middleware}
\bibfield{author}{\bibinfo{person}{Paulo Silva}, \bibinfo{person}{Daniel
  Fireman}, {and} \bibinfo{person}{Thiago Silva}.}
  \bibinfo{year}{2020}\natexlab{}.
\newblock \showarticletitle{Prebaking Functions to Warm the Serverless Cold
  Start}.
\newblock
\urldef\tempurl%
\url{https://doi.org/10.1145/3423211.3425682}
\showDOI{\tempurl}


\bibitem[\protect\citeauthoryear{Yeung}{Yeung}{[n. d.]}]%
        {Top-4-Serverless}
\bibfield{author}{\bibinfo{person}{Andy Yeung}.} \bibinfo{year}{[n.
  d.]}\natexlab{}.
\newblock \bibinfo{title}{Top 4 Serverless Computing Platforms in 2021}.
\newblock
  \bibinfo{howpublished}{https://www.loginradius.com/blog/async/serverless-overview/}.
\newblock
\newblock
\shownote{Online; Accessed: 2021-05-08.}


\end{thebibliography}

\appendix

\end{document}